\documentclass[aps,prl,twocolumn,superscriptaddress]{revtex4-2}
\usepackage{graphicx,amssymb,amsfonts,amsmath,chemarr,color,commath,braket,hyperref}

\newcommand{\beginsupplement}{%
\setcounter{page}{1}
        \setcounter{equation}{0}
        \renewcommand{\theequation}{S\arabic{equation}}%
        \setcounter{table}{0}
        \renewcommand{\thetable}{S\arabic{table}}%
        \setcounter{figure}{0}
        \renewcommand{\thefigure}{S\arabic{figure}}%
     }

\begin{document}

\title{Role of signal degradation in directional chemosensing}

\author{Ryan LeFebre}
\affiliation{Department of Physics and Astronomy, University of Pittsburgh, Pittsburgh, Pennsylvania 15260, USA}

\author{Joseph A.\ Landsittel}
\affiliation{Department of Physics and Astronomy, University of Pittsburgh, Pittsburgh, Pennsylvania 15260, USA}
\affiliation{Department of Mathematics, University of Pittsburgh, Pittsburgh, Pennsylvania 15260, USA}

\author{David E.\ Stone}
%\email{dstone@uic.edu}
\affiliation{Department of Biological Sciences, University of Illinois at Chicago, Chicago, IL 60607, USA}

\author{Andrew Mugler}
\email{andrew.mugler@pitt.edu}
\affiliation{Department of Physics and Astronomy, University of Pittsburgh, Pittsburgh, Pennsylvania 15260, USA}

\begin{abstract}
Directional chemosensing is ubiquitous in cell biology, but some cells such as mating yeast paradoxically degrade the signal they aim to detect. While the data processing inequality suggests that such signal modification cannot increase the sensory information, we show using a reaction-diffusion model and an exactly solvable discrete-state reduction that it can. We identify a non-Markovian step in the information chain allowing the system to evade the data processing inequality, reflecting the nonlocal nature of diffusion. Our results apply to any sensory system in which degradation couples to diffusion. Experimental data suggest that mating yeast operate in the beneficial regime where degradation improves sensing.
\end{abstract}

\maketitle

% Introduction
Cells actively alter their environment by secreting chemical factors. In many cases, environmental alteration by secretion of a degrading factor or by other means allows cells to form their own directional gradients out of uniform chemical backgrounds \cite{tweedy2016self}. Examples include epithelial cell migration \cite{scherber2012epithelial}, lymphocyte targeting \cite{schwab2007finding}, embryogenesis \cite{dona2013directional}, metastatic invasion \cite{muinonen2014melanoma, scherber2012epithelial}, and chemotactic bacteria \cite{fu2018spatial, cremer2019chemotaxis}. On the other hand, some cases are known in which a gradient is already present, and yet cells secrete a degrading factor anyway. A well-studied example is the chemotropic mating response of the budding yeast, {\it Saccharomyces cerevisiae} \cite{jin2011yeast}.

Haploid budding yeast cells come in two mating types. Each type secretes an attracting pheromone that is sensed by the partner type. Paradoxically, each type also secretes an enzyme that degrades the pheromone of the opposite type. Thus, a chemical gradient that each cell can use for directional sensing is already present, and yet the cell actively degrades it. It is known that this degradation is vital for efficient mating \cite{chan1982physiological, jackson1990courtship}, but the reasons are not completely understood. It has been proposed that yeast releases this degrading enzyme to disambiguate partner locations \cite{barkai1998protease}, prevent pheromone receptor saturation \cite{jin2011yeast}, and to sharpen the gradient of the pheromone profile \cite{andrews2010detailed, jin2011yeast, lakhani2017testing}.

Here, we focus on the role of degradation in sharpening the pheromone gradient because it raises a general question about the acquisition of sensory information. In principle, sharpening a gradient is useful for directional sensing, but in this case, it is achieved by the removal of pheromone. This removal will lower the overall concentration profile, which in turn should increase the sensory noise.
%It’s hard to imagine degrading (destroying) the pheromone (signal) is favorable.
If the disadvantage of increased noise outweighs the advantage of sharpening the gradient, degrading the signal may not actually be beneficial for sensing. In fact, the data processing inequality states that information cannot be increased by locally post-processing a signal \cite{cover1999elements}, which would seem to disfavor this strategy. This brings us to the central question of this paper: Is it ever beneficial for a sensory system to destroy a signal it is trying to detect?

We investigate this question using a model that accounts for molecule secretion, diffusion, and sensing by a spherical source and spherical detector. Using a perturbative approach, and accounting for diffusive fluctuations in the concentration profile, we arrive at an analytical expression for the detector's signal-to-noise ratio. As expected, we find that degradation sharpens the gradient, but it also increases noise. Taken together, the signal-to-noise ratio increases with degradation, revealing a successful sensing strategy that is nonetheless in apparent violation of data processing inequality. 

To understand this apparent violation, we reduce the model to a set of discrete states where we can calculate the mutual information between the source and detector exactly. The reduced model pinpoints a key non-Markovian step in the sensory process. Because the data processing inequality assumes Markovianity, this finding explains how the inequality is evaded. Our analysis suggests that the nonlocal character of diffusion is responsible for the information gain due to degradation. We interpret our results in terms of yeast mating but also in terms of sensory problems in general.

\begin{figure}
    \includegraphics[width=\columnwidth]{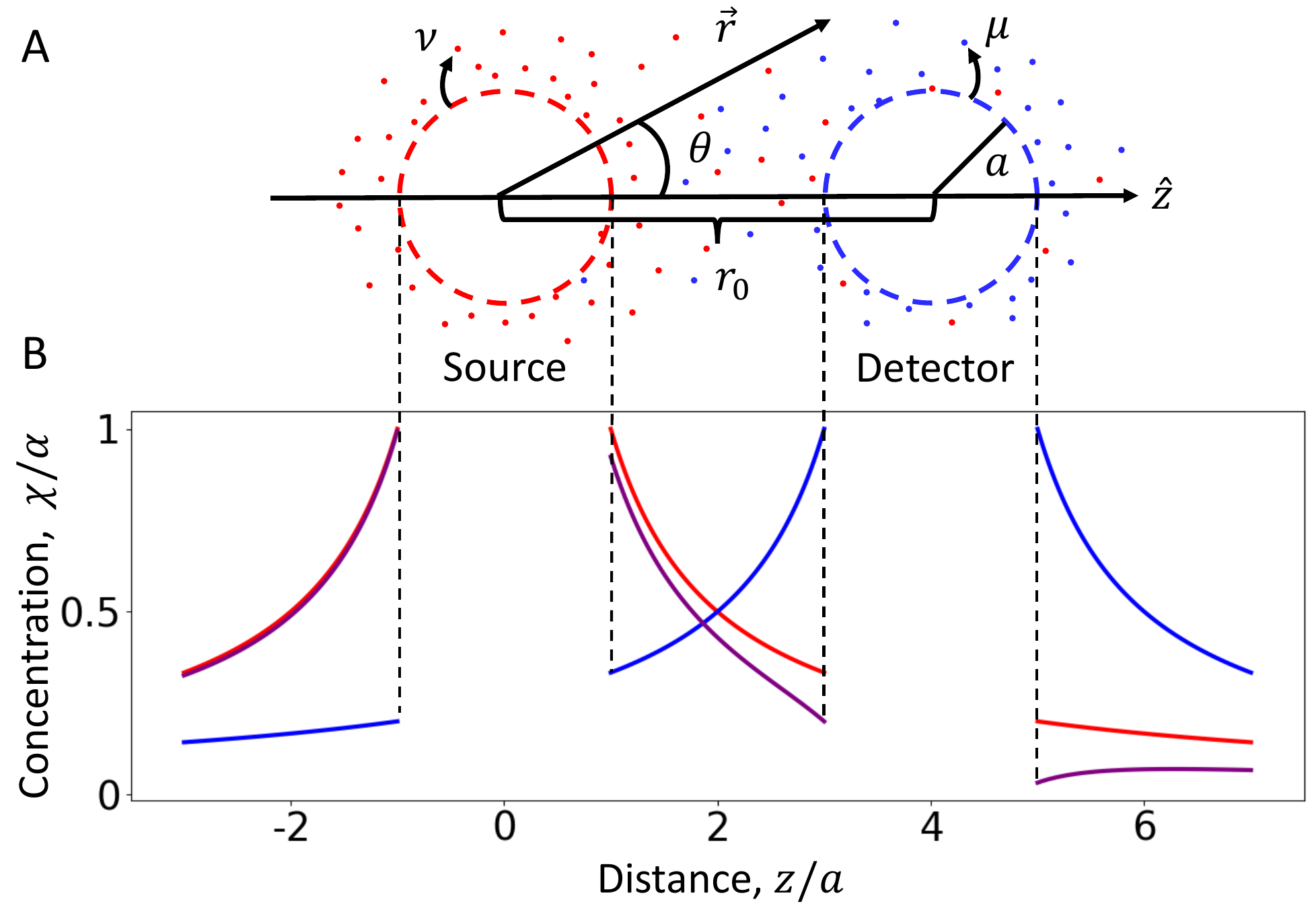}
    \caption{Model and concentration profiles. (A) A spherical source secretes a diffusing pheromone that is degraded by a diffusing enzyme secreted by a spherical detector. (B) The pheromone profile (Eq.\ \ref{chi_ss_solution}) along the $z$ axis without (red, $\varepsilon = 0$) and with (purple, $\epsilon = 0.2$) degradation for separation $\rho_0=r_0/a = 4$. The enzyme profile scales inversely with distance from the detector (blue).}
    \label{setup}
\end{figure}

% Spatiotemporal model
Consider two spheres---a source and a detector---a distance $r_0$ apart (Fig.\ \ref{setup}A). These spheres have radius $a$ and can represent whole cells themselves or, in the case of mating yeast, specific macromolecular complexes on the cell surfaces called ``gradient tracking machines'' \cite{wang2019mating}. The source releases, at rate $\nu$, an attracting pheromone with diffusion coefficient $D_c$, while the detector releases, at rate $\mu$, a degrading enzyme with diffusion coefficient $D_b$. The pheromone is degraded by the enzyme with rate $k_d$. Calling the concentrations of pheromone and enzyme $c$ and $b$, respectively, the dynamics are
\begin{align}
    \label{b_diff}
    \dot{b} &= D_b {\nabla^2} b\\
    \label{c_diff}
    \dot{c} &= D_c {\nabla^2} c - k_d b c.
\end{align}
In steady-state, Eq.\ \ref{b_diff} is solved by $b = \mu/(4 \pi D_b |\vec{r}-r_0\hat{z}|)$
%\begin{equation}
%    \label{b_ss_solution}
%    b = \frac{\mu}{4 \pi D_b |\vec{r}-r_0\hat{z}|}
%\end{equation}
in a coordinate system centered on the source (Fig.\ \ref{setup}A). Non-dimensionalizing with $\rho \equiv r/a$ and $\chi\equiv ca^3$, Eq.\ \ref{c_diff} in steady state then becomes
\begin{equation}
    \label{chi_ss_diff}
    \nabla^2 \chi = \frac{\varepsilon}{|\vec{\rho} - \rho_0 \vec{z}|} \chi,
\end{equation}
where $\varepsilon \equiv a k_d \mu/(4 \pi D_b D_c)$
%\begin{equation}
%    \label{eps}
%    \varepsilon = \frac{a k_d \mu}{4 \pi D_b D_c}
%\end{equation}
is a dimensionless parameter that reflects the strength of degradation.

Because Eq.\ \ref{chi_ss_diff} has a $\vec{\rho}$-dependent coefficient, it is not immediately solvable by linear transform methods. Therefore, we use a perturbation approach, treating $\varepsilon$ as a small parameter (an assumption we later relax when simplifying the model to a set of discrete states). Specifically, writing $\chi = \chi_0 + \varepsilon\chi_1$, the zeroth-order term satisfying $\nabla^2 \chi_0 = 0$ is $\chi_0 = \alpha/\rho$, where $\alpha \equiv \nu a^2/(4 \pi D_c)$ is a dimensionless parameter that reflects the strength of pheromone release. We solve for the first order term satisfying $\nabla^2 \chi_1 = \varepsilon\chi_0/|\vec{\rho} - \rho_0 \vec{z}|$ by expanding in spherical harmonics (see Supplemental Material). The result is
\begin{align}
\hspace{-.015in}
\chi =\ &\frac{\alpha}{\rho} - \varepsilon \alpha \sum_{\ell=0}^{\infty} P_\ell(\cos\theta)
    	\left[\frac{\ell (2\ell + 1)\rho_0 - \ell}{(\ell + 1)(2\ell + 1) \rho_0^{\ell+1} \rho^{\ell+1}} + \frac{\rho_<^\ell}{\rho_>^\ell}\right.\nonumber \\
\label{chi_ss_solution}
    & \left. -\ \frac{\rho_<^{\ell + 1}}{(2\ell + 2)\rho_>^{\ell + 1}} - \frac{1}{(2\ell + 1)(2\ell + 2)\rho_<^{\ell + 1} \rho_>^{\ell + 1}} \right],
\end{align}
where $P_\ell$ are the Legendre polynomials, and $\rho_<$ ($\rho_>$) represents the lesser (greater) of $\rho$ and $\rho_0$. Eq.\ \ref{chi_ss_solution} is plotted in Fig.\ \ref{setup}B (purple), and we see that degradation by the enzyme (blue) results in a pheromone profile that is depleted near the detector relative to that without the enzyme present (red).

\begin{figure}[b]
    \includegraphics[width=\columnwidth]{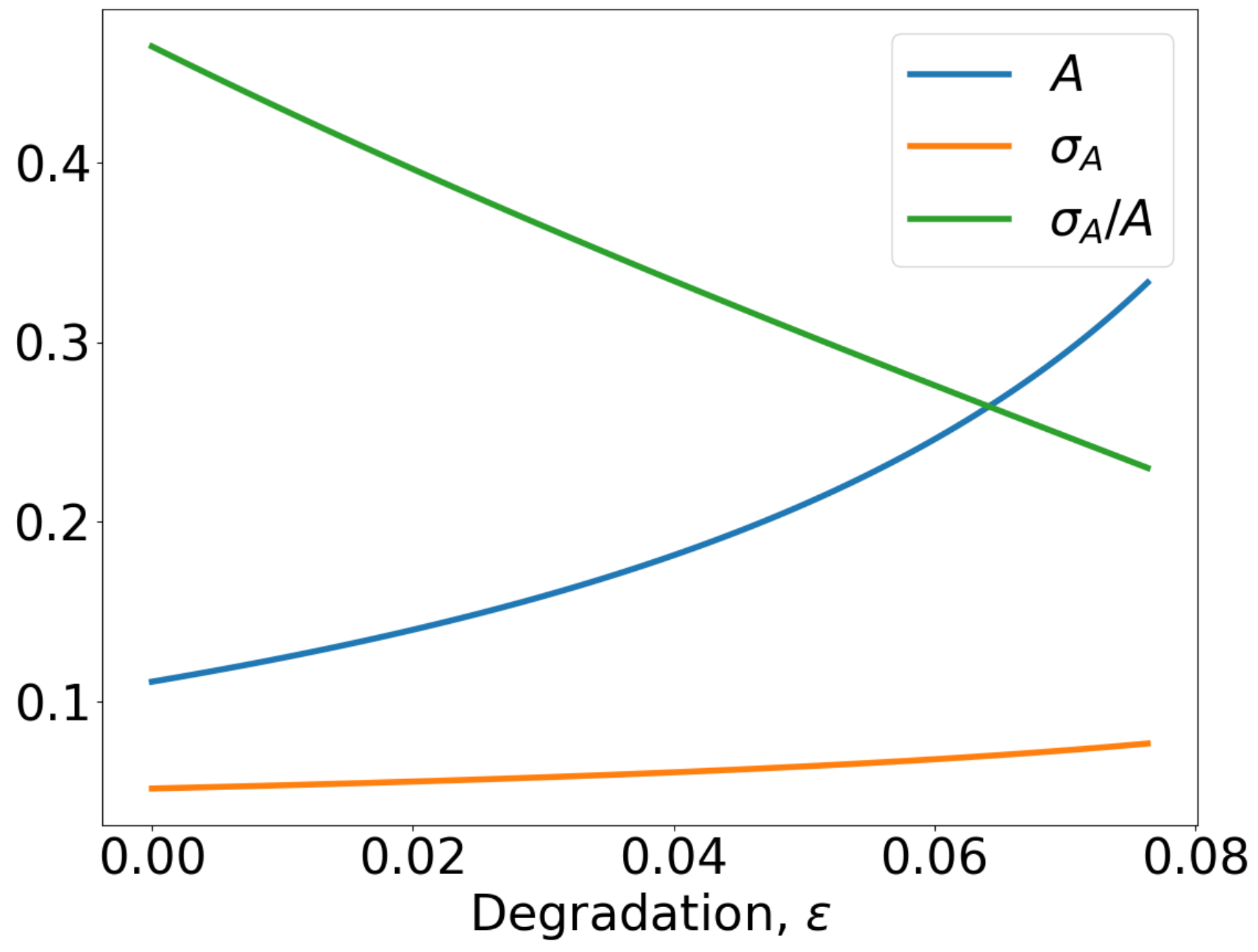}
    \caption{The anisotropy $A$ (Eq.\ \ref{anisotropy_solution}, blue) and its time-averaged variance $\sigma_A^2$ (Eq.\ \ref{noise}, orange) both increase with the dimensionless degradation parameter $\varepsilon = a k_d \mu/(4 \pi D_b D_c)$ for small $\varepsilon$. The ratio $\sigma_A/A$ (green) decreases, indicating a beneficial sensing strategy. Parameters are $\rho_0=3$, and $\nu T = 3000$.}
    \label{snr}
\end{figure}

We see in Fig.\ \ref{setup}B that degradation sharpens the pheromone gradient at the detector in the direction of the source \footnote{We also see in Fig.\ \ref{setup}B that degradation can {\it reverse} the gradient in the direction away from the source. Because the anisotropy measure we adopt in the text integrates over the entire detector surface, it accounts for the pheromone profile in all directions, including away from the source.}. Most eukaryotic cells, including yeast, do not actually measure the local gradient in a particular direction (the way that, say, motile bacteria do by moving along it \cite{berg1993random}). Rather, they compare detection events at many locations on their surface \cite{arkowitz1999responding}. The order parameter that captures this comparison is the anisotropy \cite{fancher2020precision, endres2008accuracy, Varennes:2017aa},
\begin{equation}
    \label{anisotropy_int}
    A = \frac{\int d\tilde\Omega \chi(1,\tilde\theta)\cos\tilde\theta }{\int d\tilde\Omega' \chi(1,\tilde\theta')},
\end{equation}
where $\chi(\tilde\rho,\tilde\theta)$ is Eq.\ \ref{chi_ss_solution} transformed to coordinates centered at the detector (with $\tilde\theta=0$ pointing at the source), and $d\tilde\Omega = \sin\tilde\theta d\tilde\theta d\phi$ is the corresponding solid angle element. The cosine performs the comparison, such that $A>0$ ($A<0$) corresponds to gradients toward (away from) the source. To evaluate the integrals in Eq.\ \ref{anisotropy_int}, we use a planar approximation $\chi(\tilde\rho,\tilde\theta) = c_1 + c_2\tilde z$ for the concentration profile, where the coefficients $c_1$ and $c_2$ are given by Eq.\  \ref{chi_ss_solution} at the detector surface (see Supplemental Material). Eq.\ \ref{anisotropy_int} then evaluates to
\begin{equation}
    \label{anisotropy_solution}
    A = \frac{2 + \varepsilon(G-F)(\rho_0^2 - 1)}{3[2\rho_0 - \varepsilon(F + G)(\rho_0^2 - 1)]},
\end{equation}
where $F$ and $G$ are functions of $\rho_0$ (see Supplemental Material) that satisfy $G-F>0$ when source and detector do not overlap ($\rho_0>2$). Eq.\ \ref{anisotropy_solution} is plotted in Fig.\ \ref{snr} (blue), and we see that the anisotropy $A$ increases with the degradation strength $\varepsilon$, consistent with the sharpening of the gradient.

Eq.\ \ref{anisotropy_solution} represents the detected signal but not the noise. To calculate the noise, we add Langevin terms to Eqs.\ \ref{b_diff} and \ref{c_diff} whose strengths are determined intrinsically by the parameters, and that contain the spatiotemporal correlations appropriate for diffusion \cite{fancher2020precision, Varennes:2017aa} (see Supplemental Material). Fourier transforming these equations obtains the power spectrum for $A$, whose low-frequency limit is $T\sigma^2_A$, where $\sigma^2_A$ is the variance in the long-time average of the anisotropy, and $T$ is the averaging time. The result is
\begin{equation}
\label{noise}
\sigma^2_A = \frac{2(\rho_0^2-1)}{3\nu T} \left\{1+\varepsilon\left[\frac{(F+G)(\rho_0^2-1)}{\rho_0}-\frac{3}{5}\right]\right\}.
\end{equation}
Intuition for this result can be gained from the following scaling argument. The anisotropy in Eq.\ \ref{anisotropy_int} should scale as $A\sim\Delta n/\bar{n}$, where $\Delta n = n_2-n_1$ is the front-to-back difference in the number of detected molecules, and $n = n_2+n_1$ is their sum \cite{Vennettilli2022, mugler2016limits}. The variance in the anisotropy should then scale as $\sigma_{\Delta n}^2/\bar{n}^2 \approx \sigma_n^2/\bar{n}^2$, where the second step neglects the cross-correlations between $n_2$ and $n_1$. The variance in the time-averaged anisotropy is further reduced by the number $T/\tau$ of independent measurements made in the averaging time $T$, where $\tau$ is the correlation time; hence $\sigma_A^2 \sim \sigma_n^2/(\bar{n}^2T/\tau)$. With diffusion and degradation, the statistics of the number of molecules in a given volume is Poissonian, $\sigma_n^2 = \bar{n}$, and the correlation time is set by the sum of rates set by diffusion and degradation. The diffusion rate is $D_c/a^2$, while  inspection of Eq.\ \ref{c_diff} reveals an effective degradation rate of $k_db$. Evaluating $b$ near the detector surface gives $b = \mu/(4\pi D_ba)$, and therefore a correlation time of $\tau = [D_c/a^2 + k_d\mu/(4\pi D_ba)]^{-1} = a^2/[D_c(1+\varepsilon)]$. Thus, $\sigma_A^2 \sim a^2/[\bar{n}D_c(1+\varepsilon)T]$, or recalling that $\alpha = \nu a^2/(4 \pi D_c)$, we have $\sigma_A^2 \sim \alpha/[\bar{n}\nu(1+\varepsilon)T]$. Obtaining $\bar{n}$ by integrating the concentration profile in Eq.\ \ref{chi_ss_solution} over the volume of the detector, $\bar{n} = \int d^3\tilde\rho\ \chi(\rho,\theta)$, and again using the planar approximation for $\chi$ as above, we find that this expression for $\sigma_A^2$ recovers Eq.\ \ref{noise} up to numerical factors of order unity (see Supplemental Material).
%The last equality in \ref{A_scaling2} comes from the fact that for non-extreme densities, the arrive times of molecules in an area should be independent and therefore the number of molecules in a given area should follow a Poisson distribution. 

Eq.\ \ref{noise} is plotted in Fig.\ \ref{snr} (orange), and we see that the noise $\sigma_A$ increases with the degradation strength $\varepsilon$. This result is consistent with the idea that degradation reduces the molecule number $\bar{n}$, which increases the noise as $\sigma_A \sim 1/\sqrt{\bar{n}}$ as shown above. However, we also see in Fig.\ \ref{snr} that the ratio $\sigma_A/A$ of the noise to the signal decreases with the degradation strength $\varepsilon$ (green). This result reveals that the benefit of increased signal outweighs the detriment of increased noise, such that the signal-to-noise ratio $A/\sigma_A$ still increases with degradation. Such a result would seem to be in violation of the data processing inequality, since processing (i.e., degrading) the signal has increased the sensory information (i.e., the signal-to-noise ratio).

\begin{figure}
    \includegraphics[width=\columnwidth]{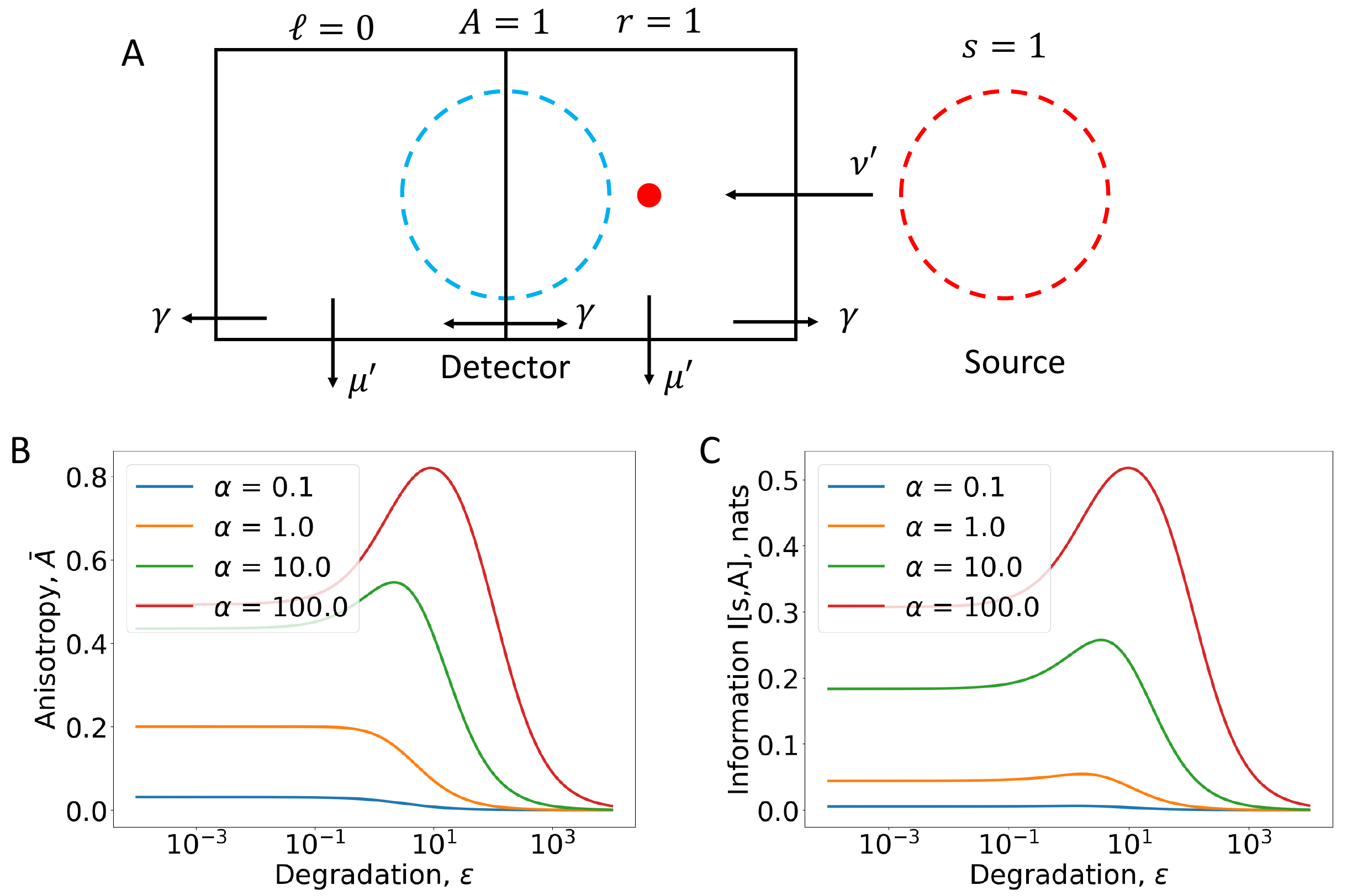}
    \caption{Discrete-state reduction. (A) Locations left and right of the detector contain either zero or one pheromone molecule. Transition rates accounting for secretion, diffusion, and degradation map to the dimensionless parameters $\alpha$ and $\varepsilon$ of the full model. (B) Average anisotropy $\bar{A}$ (right-left occupancy difference). (C) Mutual information between $A$ and source location $s$ (left or right). Both measures increase, then decrease, with degradation parameter $\varepsilon$.}
    \label{discrete}
\end{figure}

To resolve this paradox, we reduce our model to a set of discrete states, allowing us to solve for the sensory information exactly. The simplification will be considerable but will preserve all ingredients (secretion, diffusion, and degradation) and therefore the basic physics. Furthermore, it will have the added benefit that $\varepsilon$ is no longer confined to be small. Specifically, we reduce the spatial domain to two locations, one on either side of the detector, each of which may contain either zero or one pheromone molecule (Fig.\ \ref{discrete}A). The source is on one side or the other and secretes pheromone at that location. Pheromone diffuses between locations or out of the system, and is degraded. Thus, the sensory problem is reduced to: how much information does the difference in pheromone occupancies give about the source direction (left or right)?

To answer this question, we seek the conditional probability $p_{\ell r|s}$, where the binary variables $\ell\in\{0,1\}$ and $r\in\{0,1\}$ represent the left and right pheromone occupancies, and the binary variable $s\in\{-1,1\}$ represents the left-right location of the source. Calling $\nu'$, $\gamma$, and $\mu'$ the secretion, diffusion, and degradation rates, respectively (Fig.\ \ref{discrete}A), the dynamics of $p_{\ell r|s}$ are
\begin{align}
\dot{p}_{00|1} &= (\gamma + \mu')(p_{01|1} + p_{10|1}) - \nu'p_{00|1}\nonumber\\
\dot{p}_{01|1} &= \nu'p_{00|1} + \gamma p_{10|1} + (\gamma+\mu')p_{11|1} - (2\gamma+\mu')p_{01|1}\nonumber\\
\dot{p}_{10|1} &= \gamma p_{01|1} + (\gamma+\mu')p_{11|1} - (\nu'+2\gamma+\mu')p_{10|1}\nonumber\\
\label{master}
\dot{p}_{11|1} &= \nu'p_{10|1} - 2(\gamma+\mu')p_{11|1},
\end{align}
and a similar set of equations for $s=-1$ \footnote{\label{antisym}Specifically, $p_{00|1} = p_{00|-1}$, $p_{01|1} = p_{10|-1}$, $p_{10|1} = p_{01|-1}$, and $p_{11|1} = p_{11|-1}$.}. The rates $\nu'$, $\gamma$, and $\mu'$ can be mapped to their counterparts in the unreduced model as follows. The effective pheromone degradation rate is $\mu' = k_db = k_d\mu/(4\pi D_ba)$ as reasoned above. Diffusion across the detector happens at a rate on the order of $\gamma = D_c/a^2$. We thus have $\mu'/\gamma = a k_d \mu/(4 \pi D_b D_c) = \epsilon$, naturally recovering the dimensionless degradation strength within the reduced model. The effective secretion rate $\nu'$ is reduced from $\nu$ according to the distance of the detector from the source. For convenience we define $\nu'=\nu/(4\pi)$, such that $\nu'/\gamma = \alpha$, with the understanding that in general this ratio would be inversely related to $\rho_0$.
In terms of these rate ratios, the steady state of Eq.\ \ref{master} reads
\begin{align}
\label{plrs}
%p_{00|1} &= (1+\varepsilon)^2(6+2\varepsilon+\alpha)/Z \nonumber\\
%p_{01|1} &= (1+\varepsilon)\alpha(4+2\varepsilon+\alpha)/Z \nonumber\\
%p_{10|1} &= 2(1+\varepsilon)\alpha/Z \nonumber\\
%p_{11|1} &= \alpha^2/Z,
p_{00|1} &= (1+\varepsilon)^2(6+2\varepsilon+\alpha)/Z, &
p_{10|1} &= 2(1+\varepsilon)\alpha/Z, \nonumber\\
p_{01|1} &= (1+\varepsilon)\alpha(4+2\varepsilon+\alpha)/Z, &
p_{11|1} &= \alpha^2/Z,
\end{align}
where $Z\equiv(2+2\varepsilon+\alpha)[3+2\alpha+\varepsilon(4+\varepsilon+\alpha)]$.

In the reduced model, the anisotropy is equivalent to the difference in pheromone occupancies, $A = r-\ell$. Noting the simple relationship between $p_{A|s}$ and $p_{\ell r|s}$ \footnote{Specifically, the relationship between $p_{A|s}$ and $p_{\ell r|s}$ is $p_{-1|s} = p_{10|s}$, $p_{0|s} = p_{00|s} + p_{11|s}$, and $p_{1|s} = p_{01|s}$.}, the average anisotropy $\bar{A} \equiv \langle{A|s=1}\rangle$ follows from Eq.\ \ref{plrs} as
\begin{equation}
\label{Abar}
\bar{A} = \frac{\alpha(1+\varepsilon)}{3+2\alpha+\varepsilon(4+\varepsilon+\alpha)}.
\end{equation}
Eq.\ \ref{Abar} is plotted in Fig.\ \ref{discrete}B, and we see that the anisotropy $\bar{A}$ increases with degradation strength $\varepsilon$ for small $\varepsilon$, as in the unreduced model (Fig.\ \ref{snr}, blue). For large $\varepsilon$, the anisotropy ultimately vanishes, as it must for a completely degraded signal. Thus, an optimal degradation strength $\varepsilon_*$ emerges that scales as $\varepsilon_*\sim\sqrt{\alpha}$ for large $\alpha$.

The analog of the signal-to-noise ratio is the sensory information: the mutual information \cite{shannon1948mathematical} between $s$ and $A$,
\begin{align}
    I[s;A] &= \sum_{sA} p_{A|s} p_s \log\frac{p_{A|s}}{\sum_{s'}p_{A|s'}p_{s'}} \nonumber \\
    \label{MI}
    &= \frac{2\alpha(1+\varepsilon)}{Z}\left(\beta\log\frac{2\beta}{\beta+1}+\log\frac{2}{\beta+1}\right),
\end{align}
where $\beta\equiv2+\varepsilon+\alpha/2$. Here, the second step assumes the detector has no initial knowledge of the source direction ($p_s = 1/2$), inserts $p_{A|s}$ from Eq.\ \ref{plrs}, and simplifies. Eq.\ \ref{MI} is plotted in Fig.\ \ref{discrete}C, and we see that the information, like the anisotropy, increases and then decreases with $\varepsilon$. In particular, the increase shows that the reduced model reproduces the apparent violation of the data processing inequality.

Importantly, the reduced model allows us to explain the apparent violation. The information flow in the system is as follows: the source direction informs the pheromone profile, the pheromone profile is degraded, and the degraded profile informs the anisotropy. Denoting the four-state pheromone profile $(\ell,r)$ as $q$ without degradation, and as $q'$ with degradation, this flow implies the chain $s\to q\to q'\to A$. If this chain is Markovian, i.e., $p_{sqq'A} = p_sp_{q|s}p_{q'|q}p_{A|q'}$, then the data processing inequality implies $I[s;q]\ge I[s;q']\ge I[s;A]$. In the Supplemental Material we prove that $I[s;q']\ge I[s;A]$ holds but that $I[s;q]\ge I[s;q']$ does not. The latter implies that the piece of the chain $s\to q\to q'$ is non-Markovian, meaning that $q'$ is not conditionally independent of $s$ given $q$. In other words, the degraded profile $q'$ carries more information about the source direction $s$ than is contained in the unmodified profile $q$.

Why is the chain $s\to q\to q'$ non-Markovian? The reason is that degradation, when coupled to diffusion, is not a local modification to the signal. That is, when a molecule within the profile is degraded, the rest of the profile does not remain the same. Instead, diffusion reshuffles the profile, filling in the gaps to create a new steady state. This reshuffling occurs because the steady state is non-equilibrium: flux in from secretion is balanced elsewhere by flux out from diffusion and degradation. The resulting steady state is then entirely different from, and evidently more informative than, the unmodified one, despite the fact that molecules are lost.

These insights also extend to systems that form a gradient out of a uniform background. A particularly simple example is the synthesis-diffusion-degradation model of morphogenesis \cite{wartlick2009morphogen, gregor2007stability}, in which molecules enter from one side of an embryo, diffuse, and spontaneously degrade. Without degradation, diffusion would make the profile tend toward uniform. Degradation instead makes the profile fall off away from the source. Degradation thus introduces a gradient, which provides cell nuclei their positional information, despite destroying the signal that they detect.

What are the implications of our findings for mating yeast? Mating partners are typically no more than a few cell radii away, meaning that the optimal degradation condition for nearby cells, $\varepsilon_*\approx\sqrt{\alpha}$ (Fig.\ \ref{discrete}B, C), applies. If degradation is to improve sensing, we therefore must have $\varepsilon<\sqrt{\alpha}$, or, inserting the expressions for $\varepsilon$ and $\alpha$ and ignoring factors of order unity, $\mu < (D_b/k_d)\sqrt{D_c\nu}$. Note that the source or detector radius $a$ drops out of this condition, so that it does not matter for what follows whether sensing is performed by the entire cell or a local macromolecular complex. In mating yeast, the pheromone is $\alpha$-factor, which is secreted by a source cell at a rate of $\nu \approx 1350$ molecules per second in the presence of a mating partner \cite{rogers2012molecular}. The enzyme is Bar1, which binds to $\alpha$-factor with a second-order rate of $k_d = 7.7$ $\mu$M$^{-1}$s$^{-1} = 0.013$ $\mu$m$^3$/s \cite{jones2015evolutionary}. Estimating the diffusion coefficients $D_c = 125$ $\mu$m$^2$/s and $D_b = 6$ $\mu$m$^2$/s from the molecules' weights \cite{jin2011yeast}, the condition becomes $\mu < 140\nu$. Although we are unaware of measurements of the secretion rate $\mu$ of Bar1, it is unlikely that it exceeds a hundred times that of the pheromone. Therefore, our analysis predicts that mating yeast orient toward their partners under conditions in which signal degradation helps, rather than hurts, sensory precision.

We have demonstrated that degrading a directional signal can be beneficial for detecting that signal because the advantage of sharpening the gradient outweighs the disadvantage of signal loss. We have argued that this benefit is possible, despite the implications of the data processing inequality, because diffusion makes the signal modification nonlocal. The net result is an optimal level of signal degradation: enough to amplify the directional information, but not too much to destroy the signal entirely. Comparing our findings with experimental data suggests that mating yeast operate in the beneficial regime where degradation amplifies the information. Our predictions are generic and apply to any directional sensing system in which signal degradation is employed to shape or reshape a diffusive gradient.

\begin{acknowledgments}
We thank Ming Chen, Bard Ermentrout, and Bill Bialek for helpful discussions. This work was supported by National Science Foundation grant numbers PHY-2118561 and MCB-2003415.
\end{acknowledgments}

%\bibliographystyle{unsrt}
%\bibliography{refs}

\onecolumngrid
%\newpage
\vspace{.5in}

\begin{center}
    {\bf SUPPLEMENTAL MATERIAL}
\end{center}
\beginsupplement

\subsection{Pheromone and Enzyme Concentrations}

Denoting the concentration of the degrading enzyme and attractant pheromone as $b$ and $c$ 
respectively, the equations for their diffusion and interaction are as in Eqs.\ \ref{b_diff} and \ref{c_diff} of the main text,
\begin{align}
    \label{diff_b}
    \dot{b} &= D_b {\nabla^2} b \\
    \label{diff_c}
    \dot{c} &= D_c {\nabla^2} c - k_d b c.
\end{align}
Working in steady-state, it is easy to solve the equation for the degrading enzyme while centered
on the detector:
\begin{align}
    \label{b_ss_1}
    D_b {\nabla^2} b(r,\theta,\phi) &= 0\\
    \label{ss_general_solution}
    b(r,\theta,\phi) &= \sum_{\ell=0}^{\infty} \sum_{m=-\ell}^{\ell} \left( A_{\ell m}r^\ell + B_{\ell m}r^{-\ell-1} \right) Y_{\ell m}(\theta, \phi),
\end{align}
where $Y_{\ell m}$ are spherical harmonics. Noting the spherical symmetry of the problem ($l=m=0$) and using the boundary condition of $b \rightarrow 0$ as $r \rightarrow \infty$,
Eq.\ \ref{ss_general_solution} becomes
\begin{equation}
    \label{b_ss_2}
    b = \frac{B}{r}.
\end{equation}
$B$ is to be determined from the boundary condition involving the release of the enzyme through the surface of the cell
\begin{equation}
    \label{b_bc}
    -4 \pi a^2 D_b \left( \frac{\partial b}{\partial r} \right)_{r=a} = \mu.
\end{equation}
This gives the solution to $b$
\begin{equation}
    \label{b_ss_3}
    b(r) = \frac{\mu}{4 \pi D_b r}.
\end{equation}
Shifting the coordinate system to be centered on the source, Eq.\ \ref{b_ss_3} becomes
\begin{equation}
    \label{b_ss_4}
    b(r) = \frac{\mu}{4 \pi D_b |\vec{r} - r_0 \vec{z}|}.
\end{equation}
Still working in steady-state, non-dimensionalizing with $\rho \equiv r/a$ and $\chi \equiv ca^3$, and using Eq.\ \ref{b_ss_4}, Eq.\ \ref{diff_c} becomes

\begin{equation}
    \label{chi_ss_diff}
    \nabla^2 \chi = \frac{\varepsilon}{|\vec{\rho} - \rho_0 \vec{z}|} \chi,
\end{equation}
where $\varepsilon \equiv a k_d \mu/(4 \pi D_b D_c)$
is a dimensionless parameter that reflects the strength of degradation. Equation \ref{chi_ss_diff} has a variable coefficient and therefore it is not immediately 
solvable by linear transform methods. We proceed using perturbation theory for small $\varepsilon$,
\begin{equation}
    \label{chi_expansion}
    \chi = \chi_0 + \varepsilon \chi_1 + \dots.
\end{equation} 
Plugging Eq.\ \ref{chi_expansion} into Eq.\ \ref{chi_ss_diff} and matching terms up to first order we have
\begin{align}
    \label{chi0_diff}
    \nabla^2 \chi_0 &= 0\\
    \label{chi1_diff}
    \nabla^2 \chi_1 &= \frac{\varepsilon}{|\vec{\rho} - \rho_0 \vec{z}|} \chi_0.
\end{align}
The boundary conditions for $\chi$ are as follows:
\begin{equation}
    \label{chi_bc1}
    \chi (\rho \rightarrow \infty) = 0,
\end{equation}
which gives
\begin{align}
    \label{chi0_bc1}
    \chi_0 (\rho \rightarrow \infty) &= 0,\\
    \label{chi1_bc1}
    \chi_1 (\rho \rightarrow \infty) &= 0.
\end{align}
The boundary condition at the surface of the cell is
\begin{equation}
    \label{chi_bc2}
    -4 \pi a^{-2} D_c \left( \frac{\partial \chi}{\partial \rho} \right)_{\rho=1} = \nu.
\end{equation}
Rewriting $\alpha = \nu a^2 / 4 \pi D_c$, we have
\begin{equation}
    \label{chi_bc2_2}
    \left( \frac{\partial \chi_0}{\partial \rho} \right)_{\rho = 1} + \varepsilon \left( \frac{\partial \chi_1}{\partial \rho} \right)_{\rho = 1} = -\alpha.
\end{equation}
$\chi_0$ is the concentration of the pheromone when there is no enzyme present. Therefore, Eq.\ \ref{chi_bc2_2} gives,
\begin{align}
    \label{chi0_bc2}
    \left( \frac{\partial \chi_0}{\partial \rho} \right)_{\rho = 1} &= - \alpha,\\
    \label{chi1_bc2}
    \left( \frac{\partial \chi_1}{\partial \rho} \right)_{\rho = 1} &= 0.
\end{align}
Using the boundary conditions it is easy to solve Eq.\ \ref{chi0_diff},
\begin{equation}
    \label{chi0_solution}
    \chi_0 = \frac{\alpha}{\rho}.
\end{equation}
The equation for $\chi_1$ now becomes
\begin{equation}
    \label{chi1_diff_2}
    \nabla^2 \chi_1 = \frac{\alpha}{|\vec{\rho}||\vec{\rho} - \rho_0 \vec{z}|}.
\end{equation}
The solution to $\chi_1$ is the sum of the homogenous $\chi_1^h$, and particular $\chi_1^p$ solutions. We will solve the particular solution
using a Green's function, 
\begin{equation}
    \label{chi1p_eqn}
    \chi_1^p = \alpha \int d^3 \rho' \frac{G(\rho, \rho')}{|\vec{\rho'}||\vec{\rho'} - \rho_0\vec{z}|}.
\end{equation}
The Green's function of the Laplace operator is 
\begin{equation}
    \label{GF_Laplace}
    G(\rho, \rho') = \frac{-1}{4 \pi |\vec{\rho} - \vec{\rho'}|} = \sum_{\ell = 0}^{\infty} \frac{-1}{4 \pi} \frac{\rho_{<}^{\ell}}{\rho_{>}^{\ell + 1}} P_{\ell}(\cos\gamma) 
\end{equation}
where $\gamma$ is the angle between the two vectors, $\rho_{>} = \textrm{max}(\rho, \rho')$, $\rho_{<} = \textrm{min}(\rho, \rho')$, and $P_{\ell}$ 
is the Legendre polynomial. The Legendre polynomial can be expanded in terms of spherical harmonics
\begin{equation}
    \label{Legendre}
    P_{\ell}(cos(\gamma)) = \frac{4 \pi }{2 \ell + 1} \sum_{m = -\ell}^{\ell} Y_{\ell m}^{*}(\theta', \phi')Y_{\ell m}(\theta,\phi).
\end{equation}
Plugging Eqs.\ \ref{GF_Laplace} and \ref{Legendre} into Eq.\ \ref{chi1p_eqn} we get
\begin{equation}
    \label{chi1p_int_1}
    \chi_1^p = -\alpha \int d^3 \rho' \frac{1}{\rho'} \left( \sum_{\ell m} \frac{\rho_{<}^{\ell}}{\rho_{>}^{\ell + 1} (2 \ell + 1)} Y_{\ell m}^{*}(\theta', \phi')Y_{\ell m}(\theta,\phi)\right) \left( \sum_{\ell'} \frac{{\rho'}_{0<}^{\ell'}}{{\rho'}_{0>}^{\ell' + 1}} \sqrt{\frac{4 \pi}{2 \ell' + 1}} Y_{\ell' 0}(\theta^{*})\right).
\end{equation}
Here, $\rho'_{0>} = \textrm{max}(\rho',\rho_0)$ and $\rho'_{0<} = \textrm{min}(\rho',\rho_0)$.
Equation \ref{chi1p_int_1} can be simplified using the orthogonality relationship of the spherical harmonics
\begin{equation}
    \label{spherical_ortho_relationship}
    \int d \Omega' Y_{\ell m}^{*} (\theta, \phi) Y_{\ell' m' } (\theta, \phi) = \delta_{\ell \ell'} \delta_{m m'}.
\end{equation}
Rearranging the integral in Eq.\ \ref{chi1p_int_1} into radial and angular parts and then using Eq.\ \ref{spherical_ortho_relationship}, Eq.\ \ref{chi1p_int_1} becomes
\begin{equation}
    \label{chi1p_int_2}
    \chi_1^p = -\alpha \sum_{\ell = 0}^{\infty} \frac{\sqrt{4 \pi}}{(2 \ell + 1)^{3/2}} Y_{\ell 0}(\theta) \int_{1}^{\infty} d\rho' \rho' \frac{\rho_{<}^{\ell}}{\rho_{>}^{\ell + 1}} \frac{{\rho'}_{0<}^{\ell}}{{\rho'}_{0>}^{\ell + 1}}.
\end{equation}
The integral in Eq.\ \ref{chi1p_int_2} depends on whether $\rho$ is greater than or less than $\rho_0$.
\begin{align}
    \label{rho_lessthan}
    \int_{1}^{\infty} d\rho' \rho' \frac{\rho_{<}^{\ell}}{\rho_{>}^{\ell + 1}} \frac{{\rho'}_{0<}^{\ell'}}{{\rho'}_{0>}^{\ell' + 1}} &= \int_{1}^{\rho} d\rho' \frac{{\rho'}^{2 \ell +1}}{(\rho \rho_0)^{\ell + 1}} + \int_{\rho}^{\rho_0} d\rho' \frac{\rho^{\ell}}{\rho_0^{\ell+1}} + \int_{\rho_0}^{\infty} d\rho' \frac{(\rho \rho_0)^{\ell}}{{\rho'}^{2 \ell + 1}} \qquad \rho < \rho_0\\
    \label{rho_greaterthan}
    \int_{1}^{\infty} d\rho' \rho' \frac{\rho_{<}^{\ell}}{\rho_{>}^{\ell + 1}} \frac{{\rho'}_{0<}^{\ell'}}{{\rho'}_{0>}^{\ell' + 1}} &= \int_{1}^{\rho_0} d\rho' \frac{{\rho'}^{2 \ell +1}}{(\rho \rho_0)^{\ell + 1}} + \int_{\rho_0}^{\rho} d\rho' \frac{\rho_0^{\ell}}{\rho^{\ell+1}} + \int_{\rho}^{\infty} d\rho' \frac{(\rho \rho_0)^{\ell}}{{\rho'}^{2 \ell + 1}} \qquad \rho > \rho_0. 
\end{align}
Solving the integrals in Eqs.\ \ref{rho_lessthan} and \ref{rho_greaterthan} and writing the solution in terms of Legendre polynomials, the particular solution is
\begin{equation}
    \label{chi1p_solution}
    \chi_1^p = - \alpha \sum_{\ell = 0}^{\infty} \frac{P_{\ell}(\cos\theta)}{2 \ell + 1} \left[ (2 \ell + 1)\left(\frac{\rho_<}{\rho_>}\right)^{\ell} - \frac{2 \ell + 1}{2 \ell + 2} \left(\frac{\rho_<}{\rho_>}\right)^{\ell + 1} - \frac{1}{(2 \ell + 2)\left(\rho_< \rho_>\right)^{\ell + 1}} \right],
\end{equation}
where $\rho_>$ and $\rho_<$ are $\textrm{max}(\rho,\rho_0)$ and $\textrm{min}(\rho,\rho_0)$ respectively.
The homogenous solution is the fundamental solution to Laplace's equation with azimuthal symmetry,
\begin{equation}
    \label{chi1h_general}
    \chi_1^h = \sum_{\ell=0}^{\infty} \left( A_{\ell}\rho^{\ell} + B_{\ell}\rho^{-\ell-1} \right) P_{\ell}(\cos\theta).
\end{equation}
Applying the boundary condition from Eq.\ \ref{chi1_bc1} means that $A_{\ell} = 0$ for all $\ell$,
\begin{equation}
    \label{chi1h_general2}
    \chi_1^h = \sum_{\ell=0}^{\infty} \left( \frac{ B_{\ell}}{\rho^{\ell+1}} \right) P_{\ell}(\cos\theta).
\end{equation}
The $B_{\ell}$ can be solved for using the remaining boundary condition, Eq.\ \ref{chi1_bc2},
\begin{align}
    \label{chi1_bc2_2}
    0 &= -\sum_{\ell = 0}^{\infty} B_{\ell} (\ell + 1) P_{\ell} (\cos\theta) - \alpha \sum_{\ell=0}^{\infty} \frac{P_{\ell}(\cos\theta)}{2\ell + 1} \left[ \frac{\ell}{\rho_0^{\ell}} \left( 2 \ell + 1 - \frac{1}{\rho_0} \right) \right]\\
    \label{B_l}
    B_{\ell} &= \frac{-\alpha \ell \left( 2 \ell + 1 - 1/\rho_0 \right)}{(\ell + 1)(2 \ell + 1) \rho_0^{\ell}}.
\end{align}
Putting it all together we have,
\begin{equation}
    \label{chi}
    \chi = \frac{\alpha}{\rho} - \varepsilon \left[ \alpha \sum_{\ell}^{\infty} P_{\ell} (\cos\theta) \left( \frac{\ell \left(2 \ell + 1 - 1/\rho_0 \right)}{(\ell + 1)(2 \ell + 1)\rho_0^{\ell}\rho^{\ell+1}} + \left(\frac{\rho_<}{\rho_>}\right)^{\ell} - \frac{1}{(2 \ell + 2)} \left(\frac{\rho_<}{\rho_>}\right)^{\ell + 1} - \frac{1}{(2 \ell +1)(2 \ell + 2)(\rho_< \rho_>)^{\ell + 1}} \right) \right],
\end{equation}
as in Eq.\ \ref{chi_ss_solution} of the main text.

\subsection{Anisotropy}

To solve for the anisotropy, $A$, it is beneficial to shift the coordinate system to be centered on the detector. In this system, 
$\tilde{\rho}$ and $\tilde{\theta}$ are the new coordinates and $\tilde{\theta} = 0$ points in the direction of the source. 
The anisotropy is then defined as
\begin{equation}
    \label{anisotropy_definition}
    A = \frac{\int d\tilde{\Omega} \chi(1,\tilde{\theta})\cos\tilde{\theta}}{\int d\tilde{\Omega}' \chi(1,\tilde{\theta}')},
\end{equation}
as in Eq.\ \ref{anisotropy_int} of the main text.
To evaluate the integrals in Eq.\ \ref{anisotropy_definition} we use a planar approximation for $\chi(\tilde{\rho},\tilde{\theta})$,
\begin{equation}
    \label{chi_planar}
    \chi(\tilde{\rho},\tilde{\theta}) = c_1 + c_2 \tilde{z} = c_1 + c_2 \tilde{\rho} \cos \tilde{\theta},
\end{equation}
where $c_1$ and $c_2$ are given by Eq.\ \ref{chi} at the surface of the detector,
\begin{align}
    \label{coef_plus}
    \chi(\tilde{\rho}=1,\tilde{\theta}=0) &= \chi(\rho=\rho_0-1,\theta=0) = c_1 + c_2,\\
    \label{coef_minus}
    \chi(\tilde{\rho}=1,\tilde{\theta}=\pi) &= \chi(\rho=\rho_0+1,\theta=0) = c_1 - c_2.
\end{align}
Rearranging and solving for $c_1$ and $c_2$ we have
\begin{align}
    \label{c1_1}
    c_1 &= \frac{1}{2}(\chi(\rho=\rho_0-1,\theta=0) + \chi(\rho=\rho_0+1,\theta=0)),\\
    \label{c2_1}
    c_2 &= \frac{1}{2}(\chi(\rho=\rho_0-1,\theta=0) - \chi(\rho=\rho_0+1,\theta=0)).
\end{align}
Plugging in $\rho = \rho_0 - 1$, $\rho = \rho_0 + 1$, and $\theta = 0$ into Eq.\ \ref{chi}, we have
\begin{align}
    \label{chi_rho0minus}
    \chi(\rho_0-1,0) &= \frac{\alpha}{\rho_0 - 1} - \varepsilon F,\\
    \label{chi_rho0plus}
    \chi(\rho_0+1,0) &= \frac{\alpha}{\rho_0 + 1} - \varepsilon G,
\end{align}
where we have defined $F$ and $G$ as 
\begin{align}
    \label{F_1}
    F &= \alpha \sum_{\ell}^{\infty} \left( \frac{\ell \left(2 \ell + 1 - 1/\rho_0 \right)}{(\ell + 1)(2 \ell + 1)\rho_0^{\ell}(\rho_0-1)^{\ell+1}} + \left(\frac{\rho_0-1}{\rho_0}\right)^{\ell} - \frac{1}{(2 \ell + 2)} \left(\frac{\rho_0-1}{\rho_0}\right)^{\ell + 1} - \frac{1}{(2 \ell +1)(2 \ell + 2)((\rho_0-1) \rho_0)^{\ell + 1}} \right),\\
    \label{G_1}
    G &= \alpha \sum_{\ell}^{\infty} \left( \frac{\ell \left(2 \ell + 1 - 1/\rho_0 \right)}{(\ell + 1)(2 \ell + 1)\rho_0^{\ell}(\rho_0+1)^{\ell+1}} + \left(\frac{\rho_0}{\rho_0+1}\right)^{\ell} - \frac{1}{(2 \ell + 2)} \left(\frac{\rho_0}{\rho_0+1}\right)^{\ell + 1} - \frac{1}{(2 \ell +1)(2 \ell + 2)(\rho_0 (\rho_0+1))^{\ell + 1}} \right).
\end{align}
Performing the infinite series gives
\begin{align}
    \label{F_2}
    F &= \alpha \left[ \frac{\rho_0^2(\rho_0-1)}{\rho_0(\rho_0-1)-1} - \frac{1}{2}\text{log}\rho_0 + \frac{1}{2}\text{log}\left( \frac{\rho_0(\rho_0-1)-1}{\rho_0(\rho_0-1)} \right) \right],\\
    \label{G_2}
    G &= \alpha \left[ \frac{(\rho_0+1)(\rho_0^2+\rho_0-1)+\rho_0}{\rho_0^2+\rho_0-1} - \frac{1}{2}\text{log}(\rho_0+1) + \frac{1}{2}\text{log}\left( \frac{\rho_0^2+\rho_0-1}{\rho_0(\rho_0+1)} \right) \right].
\end{align}
The coefficients $c_1$ and $c_2$ now take the form of 
\begin{align}
    \label{c1_2}
    c_1 &= \frac{\alpha\rho_0}{\rho_0^2 - 1} - \varepsilon \frac{1}{2} (F + G),\\
    \label{c2_2}
    c_2 &= \frac{\alpha}{\rho_0^2-1} - \varepsilon \frac{1}{2} (F - G).
\end{align}
The integrals in the anisotropy, Eq.\ \ref{anisotropy_definition}, can now be evaluated, giving
\begin{equation}
    \label{A}
    A = \frac{2+\varepsilon(G-F)(\rho_0^2-1)}{3[2\rho_0-\varepsilon(F+G)(\rho_0^2-1)]},
\end{equation}
as in Eq.\ \ref{anisotropy_solution} of the main text.

\subsection{Noise: Langevin Analysis}

Equation \ref{A} in the previous section represents the detected signal of the steady-state concentration. To calculate the noise, 
we add Langevin terms to Eqs.\ \ref{diff_b} and \ref{diff_c} and make use of the Wiener-Khinchin theorem. 
The Wiener-Khinchin theorem shows that the autocorrelation and power spectral density of a signal form a Fourier pair. It can be 
used further to show that the time averaged correlation function of a signal, $C_T(t)$, is approximately equal to its power spectrum density, 
$S(\omega)$, at low frequency divided by the averaging time, $T$
\begin{equation}
    \label{WKt}
    C_T(0) = \frac{S(0)}{T}.
\end{equation}
Therefore, the time-averaged variance of the anisotropy is 
\begin{equation}
    \label{ta_A_var}
    \sigma_A^2 = \frac{S_A(\omega = 0)}{T} = \frac{1}{T} \int d \omega' \langle \tilde{\delta A}^{*}(\omega) \tilde{\delta A}(\omega=0)\rangle.
\end{equation}
$\tilde{\delta A}$ represents the fluctuations in the Fourier transform of $A(t)$ and $*$ denotes the complex conjugate. The fluctuations in 
$A(t)$ in time are by definition
\begin{equation}
    \label{deltaAt}
    \delta A (t) = \frac{\int d \Omega \delta c (a,\theta,t) \cos\theta}{\int d \Omega \bar{c} (a,\theta)}, 
\end{equation}
where $\delta c$ represents the fluctuations in the pheromone concentration and $\bar{c}$ is the time average of the pheromone concentration, given by Eqs.\ \ref{chi_planar}, \ref{c1_2}, and \ref{c2_2} (dropping the tildes for convenience).  
Denoting the denominator of Eq.\ \ref{deltaAt} as $Z$, we have
\begin{equation}
    \label{Z}
    Z = \int d \Omega \bar{c} (a,\theta) = (4 \pi) \frac{2\alpha\rho_0 - \varepsilon (F+G) (\rho_0^2 - 1)}{2a^3(\rho_0^2-1)}. 
\end{equation} 
Taking the Fourier transform of Eq.\ \ref{deltaAt} gives
\begin{equation}
    \label{deltaAw}
    \tilde{\delta A}(\omega) = \frac{1}{Z} \int d \Omega \int \frac{d^3k}{(2\pi)^3} \tilde{\delta c}(\vec{k},\omega) e^{i(a,\Omega)\cdot\vec{k}} \cos\theta.
\end{equation}
Using the plane-expansion
\begin{equation}
    \label{plane_wave}
    e^{i(a,\Omega)\cdot\vec{k}} \cos\theta = 4 \pi \sum_{\ell,m} i^{\ell} j_{\ell} (ak) Y_{\ell,m}(\Omega) Y_{\ell,m}(\hat{k}) \sqrt{\frac{4 \pi}{3}} Y_{1,0}(\Omega),
\end{equation}
where $i$ is the imaginary number and $j_\ell$ are the spherical Bessel functions. 
Using the orthogonality properties of the spherical harmonics (Eq.\ \ref{spherical_ortho_relationship}), Eq.\ \ref{deltaAw} becomes 
\begin{equation}
    \label{detlaAw2}
    \tilde{\delta A}(\omega) = \frac{1}{Z} \int \frac{d^3k}{(2\pi)^3} \tilde{\delta c}(\vec{k},\omega) 4 \pi i \cos\theta_k j_1(ak),
\end{equation}
where $\theta_k$ is polar angle in Fourier space. Taking the complex conjugate and averaging gives
\begin{equation}
    \label{AA_ave}
    \langle \tilde{\delta A}^{*}(\omega') \tilde{\delta A}(\omega) \rangle = \frac{4}{Z^2 (2 \pi)^4} \int d^3k' \cos\theta_{k'} j_{1}(ak') \int d^3k \cos\theta_k j_1 (ak) \langle \tilde{\delta c}^{*}(k',\omega') \tilde{\delta c} (k,\omega) \rangle. 
\end{equation}
To solve for $\langle \tilde{\delta c}^{*}(k',\omega') \tilde{\delta c} (k,\omega) \rangle$ we go back to our original 
PDE system, with the coordinate system centered on the detector, and with the boundary conditions built in:
\begin{equation}
    \label{diff_b_noise}
    \dot{b} = D_b \nabla^2 b + \mu \delta^3(\vec{x}) + \eta_b
\end{equation}
\begin{equation}
    \label{diff_c_noise}
    \dot{c} = D_c \nabla^2 c + \nu \delta^3(\vec{x}-r_0\vec{z}) + \eta_c
\end{equation}
The statistics of the noise terms are
\begin{equation}
    \label{etab_stats}
    \langle \eta_b(\vec{x},t) \eta_b(\vec{x'},t') \rangle = 2 D_b \delta(t-t') \vec{\nabla}_x \cdot \vec{\nabla}_{x'} [\bar{b}(\vec{x})\delta^3(\vec{x}-\vec{x'})]
\end{equation}
and 
\begin{equation}
    \label{etac_stats}
    \langle \eta_c(\vec{x},t) \eta_c(\vec{x'},t') \rangle = 2 D_c \delta(t-t') \vec{\nabla}_x \cdot \vec{\nabla}_{x'} [\bar{c}(\vec{x})\delta^3(\vec{x}-\vec{x'})] + k_d \bar{b}(\vec{x}) \bar{c}(\vec{x}) \delta(t-t') \delta^3(\vec{x}-\vec{x'}).
\end{equation}
Here, $\bar{b}$ and $\bar{c}$ are the time averages of $b(\vec{x},t)$ and $c(\vec{x},t)$ respectively. Letting $b(\vec{x},t) =\bar{b} + \delta b(\vec{x},t)$ and $c(\vec{x},t) =\bar{c} + \delta c(\vec{x},t)$, Eqs.\ \ref{diff_b_noise} and \ref{diff_c_noise} become 
\begin{equation}
    \label{diff_deltab}
    \dot{\delta b} = D_b \nabla^2 \delta b + \eta_b
\end{equation}
\begin{equation}
    \label{diff_deltac}
    \dot{\delta c} = \nabla^2 \delta c - k_d\bar{b}\bar{c} - k_d\bar{b}\delta c - k_d b \bar{c} + \eta_c.
\end{equation}
In Eq.\ \ref{diff_deltac} we have neglected second order terms in time. Taking the Fourier transform of both sides of Eq.\ \ref{diff_deltac} and rearranging we have
\begin{equation}
    \label{deltacw}
    \tilde{\delta c}(\vec{k},\omega) = \frac{\eta_c - k_d \bar{c} \tilde{\delta b}}{D_b k^2 + k_d \bar{b} - i \omega}.
\end{equation}
Taking the complex conjugate of Eq.\ \ref{diff_deltac} and averaging gives
\begin{equation}
    \label{cc_ave}
    \langle \tilde{\delta c}^{*}(\vec{k'},\omega') \tilde{\delta c}(\vec{k},\omega) \rangle = \frac{\langle \eta_c^* \eta_c \rangle + (k_d \bar(c))^2 \langle \tilde{\delta b}^{*} \tilde{\delta b} \rangle }{(D_ck'^2 + k_d \bar{b} + i \omega')(D_c k^2 + k_d \bar{b} - i \omega)}.
\end{equation}
A similar procedure can be followed for $\delta b$ giving

\begin{equation}
    \label{bb_ave}
    \langle  \tilde{\delta b}(\vec{k'},\omega') \tilde{\delta b}(\vec{k},\omega) \rangle = \frac{\langle  \tilde{\eta_b}^{*} \tilde{\eta_b} \rangle}{(D_b k'^2 + i \omega')(D_b k^2 - i \omega)}.
\end{equation}
Plugging Eqs.\ \ref{bb_ave}, \ref{etab_stats}, and \ref{etac_stats} into Eq.\ \ref{cc_ave} gives
\begin{equation}
    \label{cc_ave_2}
    \langle \tilde{\delta c}^{*}(\vec{k'},\omega') \tilde{\delta c}(\vec{k},\omega) \rangle = \frac{(4\pi)^4 \delta^3(\vec{k'}-\vec{k})\left[ 2D_c \bar{c} k^2 + k\bar{b}\bar{c} + 2(k_d\bar{c})^2D_b\bar{b}k^2/(D_b^2k^4 + \omega^2) \right]}{(D_ck^2+(k_d\bar{b})^2)+\omega^2}.
\end{equation}
We can now solve for $\sigma_A^2$ by plugging the above equation into Eq.\ \ref{AA_ave},
\begin{equation}
    \label{ta_A_var_2}
    \sigma_A^2 = \frac{4}{Z^2T} \int_{\-\infty}^{\infty} dk j_1^2(ka) \left[ \frac{2D_bD_c\bar{c}k^4 + k_d\bar{b}\bar{c}k^2  + 2 (k_d \bar{c})^2 \bar{b}}{D_b(D_ck^2 + k_d\bar{b})^2} \right]. 
\end{equation}
Defining $u \equiv ka$, using $\bar{b} = \mu / 4 \pi D_b a$, and recalling that $\varepsilon = k_d \mu a / 4 \pi D_b D_c$, Eq.\ \ref{ta_A_var_2} becomes
\begin{equation}
    \label{ta_A_var_3}
    \sigma_A^2 = \frac{4\bar{c}}{Z^2aD_cT} \int_{-\infty}^{\infty} du j_1^2(u) \left[ \frac{2u^4 + \varepsilon u^2 + 2f\varepsilon^2}{(u^2 + \varepsilon)^2} \right]
\end{equation}
where $f = \nu / \rho_0 \mu$. The above integral can be broken up into three separate integrals which in turn can be solved by integration by parts and contour integrations.
\begin{equation}
    \label{int_1}
    I_1 = \int_{-\infty}^{\infty} du j_1^2(u) \frac{2u^4}{(u+i\sqrt{\varepsilon})^2(u-i\sqrt{\varepsilon})^2} = \frac{\pi[1 + \varepsilon - e^{-2\varepsilon^{1/2}}(1 + 2 \varepsilon^{1/2} + 3\varepsilon + 2 \varepsilon^{3/2})]}{2 \varepsilon^{3/2}} 
\end{equation}    
\begin{equation}
    \label{int_2}
    I_ 2 = \int_{-\infty}^{\infty} du j_1^2(u) \frac{\varepsilon u^2}{(u+i\sqrt{\varepsilon})^2(u-i\sqrt{\varepsilon})^2} = \frac{\pi[-3 + \varepsilon + e^{-2\varepsilon^{1/2}}(3 + 6 \varepsilon^{1/2} + 5\varepsilon + 2 \varepsilon^{3/2})]}{4 \varepsilon^{3/2}} 
\end{equation} 
\begin{equation}
    \label{int_3}
    I_3 = \int_{-\infty}^{\infty} du j_1^2(u) \frac{2f\varepsilon}{(u+i\sqrt{\varepsilon})^2(u-i\sqrt{\varepsilon})^2} = \frac{\pi f[15 - 9 \varepsilon + 4 \varepsilon^{3/2} - e^{-2\varepsilon^{1/2}}(15 + 30 \varepsilon^{1/2} + 21\varepsilon + 6 \varepsilon^{3/2})]}{6 \varepsilon^{3/2}}. 
\end{equation} 
To first order in $\varepsilon$, 
\begin{equation}
    \label{Is}
    I_1 + I_2 + I_3 \approx \pi\left( \frac{2}{3} - \varepsilon \frac{2}{5} \right).
\end{equation}
Plugging in Eqs.\ \ref{Z} and \ref{Is} into Eq.\ \ref{ta_A_var_3} and using $\bar{c} = c_1/a^3 = \nu \rho_0 / 4 \pi D_c a (\rho_0^2 - 1)$, we have the variance in the time-average anisotropy,
\begin{equation}
    \label{ta_A_var_4}
    \sigma_A^2 = \frac{2(\rho_0^2-1)}{3\nu T}\left[ 1 + \varepsilon \left[ (F+G)\left( \frac{\rho_0^2 - 1}{\rho_0} \right) - \frac{3}{5}\right] \right],
\end{equation}
as in Eq.\ \ref{noise} of the main text.

\subsection{Noise: Scaling Argument}

According to the scaling argument of the main text, the variance of the long-time average of the anisotropy is
\begin{equation}
    \label{A_var_2}
    \sigma_A^2 \sim \frac{a^2}{D_c T (1 + \varepsilon) \bar{n}},
\end{equation}
where
\begin{equation}
    \label{nbar_def}
    \bar{n} = \int d^3\tilde{\rho} \chi(\tilde{\rho},\tilde{\theta})
\end{equation}
is the number of pheromone molecules within the volume of the detector. Eq.\ \ref{nbar_def}
can be solved using Eq.\ \ref{chi_planar}
\begin{equation}
    \label{nbar}
    \bar{n} = \frac{4 \pi c_1}{3} = \frac{4 \pi (2 \alpha \rho_0 - \varepsilon (F+G)(\rho_0^2 -1))}{6(\rho_0^2-1)}.
\end{equation}
Plugging Eq.\ \ref{nbar} into Eq.\ \ref{A_var_2} and using $\alpha = \nu a^2 / 4 \pi D_c$ we have
\begin{equation}
    \label{A_var_3}
    \sigma_A^2 \sim \frac{3(\rho_0^2-1)}{\nu T \rho_0 (1+\varepsilon)\left[ 1 - \varepsilon (F + G) (\rho_0^2-1)/(2\rho_0) \right]}.
\end{equation}
To first order in $\varepsilon$, the variance is 
\begin{equation}
    \label{A_var_4}
    \sigma_A^2 \sim \frac{3(\rho_0^2-1)}{\nu T \rho_0} \left[ 1 + \varepsilon \left( \frac{1}{2}(F+G)\left(\frac{\rho_0^2 -1 }{\rho_0}\right) - 1\right) \right].
\end{equation}
Comparing Eq.\ \ref{A_var_4} with Eq.\ \ref{ta_A_var_4}, we see that the two expressions agree apart from numerical factors, as stated in the main text.

\subsection{Information Bounds}

Here we prove that the reduced model satisfies the bound $I[s;q']\ge I[s;A]$, but not the bound $I[s;q]\ge I[s;q']$. Here $s\in\{-1,1\}$ is the source location, $q = (\ell,r)\in\{(0,0),(0,1),(1,0),(1,1)\}$ are the pheromone occupancies without degradation, $q'$ are the pheromone occupancies with degradation, and $A \in\{-1,0,1\}$ is the anisotropy.

To prove the first bound, we recognize that the information between source and degraded profile is
\begin{align}
I[s;q'] &= \sum_{s\ell r}p_{\ell r|s}p_s\log\frac{p_{\ell r|s}}{\sum_{s'}p_{\ell r|s'}p_{s'}}\\
\label{Isq'}
&= \frac{1}{2}\sum_{\ell r}p_{\ell r|-1}\log\frac{2p_{\ell r|-1}}{p_{\ell r|-1}+p_{\ell r|1}}
	+ \frac{1}{2}\sum_{\ell r}p_{\ell r|1}\log\frac{2p_{\ell r|1}}{p_{\ell r|-1}+p_{\ell r|1}},
\end{align}
where the second step applies $p_s = 1/2$. We introduce the shorthand
\begin{align}
p_{00|1} &\equiv a, & p_{00|-1} &= a, \nonumber \\
p_{01|1} &\equiv b, & p_{01|-1} &= -b, \nonumber \\
p_{10|1} &\equiv c, & p_{10|-1} &= -c, \nonumber \\
\label{abcd}
p_{11|1} &\equiv d, & p_{11|-1} &= d,
\end{align}
where the second column follows from the antisymmetry of the solution upon swapping the source location. Inserting Eq.\ \ref{abcd} into Eq.\ \ref{Isq'} and simplifying obtains
\begin{equation}
\label{Isq'2}
I[s;q'] = b\log\frac{2b}{b+c} + c\log\frac{2c}{b+c}.
\end{equation}
The information between source and anisotropy is
\begin{align}
I[s;A] &= \sum_{sA}\tilde{p}_{A|s}p_s\log\frac{\tilde{p}_{A|s}}{\sum_{s'}\tilde{p}_{A|s'}p_{s'}}\\
\label{IsA}
&= \frac{1}{2}\sum_A\tilde{p}_{A|-1}\log\frac{2\tilde{p}_{A|-1}}{\tilde{p}_{A|-1}+\tilde{p}_{A|1}}
	+ \frac{1}{2}\sum_A\tilde{p}_{A|1}\log\frac{2\tilde{p}_{A|1}}{\tilde{p}_{A|-1}+\tilde{p}_{A|1}},
\end{align}
where for clarity we have used a tilde to distinguish the anisotropy distribution $\tilde{p}_{A|s}$ from the profile distribution $p_{\ell r|s}$. Given the relationship $A = \ell-r$, we have
\begin{align}
\tilde{p}_{-1|1} &= p_{01|1} = b, & \tilde{p}_{-1|-1} &= p_{01|-1} = -b, \nonumber \\
\tilde{p}_{0|1} &= p_{00|1} + p_{11|1} = a+d,
	& \tilde{p}_{0|-1} &= p_{00|-1} + p_{11|-1} = a+d, \nonumber \\
\label{abcd2}
\tilde{p}_{1|1} &= p_{10|1} = c, & \tilde{p}_{1|-1} &= p_{10|-1} = -c.
\end{align}
Inserting Eq.\ \ref{abcd2} into Eq.\ \ref{IsA} and simplifying obtains
\begin{equation}
\label{IsA2}
I[s;A] = b\log\frac{2b}{b+c} + c\log\frac{2c}{b+c}.
\end{equation}
Because Eqs.\ \ref{Isq'2} and \ref{IsA2} are identical, we conclude that $I[s;q'] = I[s;A]$. That is, the reduced model satisfies the first bound $I[s;q']\ge I[s;A]$ by equality.

The violation of the second bound can now be shown by counterexample. We see from Fig.\ \ref{discrete}C that $I[s;A]$ with degradation ($\epsilon > 0$) can be larger than without degradation ($\epsilon = 0$). Because we just found that $I[s;A] = I[s;q']$, we therefore conclude that $I[s;q']$ with degradation can be larger than without degradation---which we denoted $I[s;q]$. This example thus shows that the second bound $I[s;q]\ge I[s;q']$ does not generally hold.

\end{document}